# Effect of support on the vanadyl oxygen abstraction in supported vanadia

Viktor Kovalskii[1], Igor Zilberg[*1,2]

*[1]Boreskov Institute of Catalysis, 630090 Novosibirsk, Russia*

*[2]Novosibirsk State University, 630090 Novosibirsk, Russia*

Supported vanadia catalysts are modeled within the cluster DFT approach to get an insight into the mechanism in which the support affects the activity of vanadia in the oxidation processes. The energy of the V=O group dissociation chosen as a descriptor of the oxidation activity is estimated using two aligned divanadate $V_2O_3(OH)_4$ particles at various distances. Separation between particles allows to imitate (i) the various supporting materials (e.g. $TiO_2$, $SiO_2$, etc.), and (ii) the coverage of vanadia on a particular support. A substantial compensation of the energy loss upon the vanadyl oxygen abstraction via bonding to the vanadyl oxygen of neighboring vanadate particle has been predicted. On account of such compensation the overall energy of the V=O dissociation reaches its minimal value of 36 kcal/mol (dropping from maximum of 142 kcal/mol) at small separation of 3 Å between dimers when the nearest vanadyl oxygen occupies a bridge V-O-V position. For dimers separated by about 4 Å, the dissociation energy achieves its maximal value for isolated dimers of about 143 kcal/mol. Thus, these findings allows one to conclude that the oxygen mobility is a result of a compensation effect in a chain-like bonding between neighboring vanadyl groups on the surface of support.

# I. Introduction

Supported vanadia catalysts, particularly $V_2O_5/TiO_2$, are widely used in the reactions of selective oxidation such as, for example, selective oxidation of o-xylene to phthalic anhydride [1,2], oxidative ammonolysis of alkyl-substituted aromatic compounds [3], selective catalytic reduction of $NO_x$ with $NH_3$ [4] and controlling the $SO_2$ to $SO_3$ oxidation in the selective catalytic reduction [5–7]. In addition to industrial oxidation reactions supported vanadia catalysts are used for dehydration alkanes to olefins [8], oxidation butane to maleinic anhydride [9–11], oxidation of penthane to maleinic anhydride [12], selective oxidation of methanol to formaldehyde [13] or methylformate [14].

---

[*] Corresponding author, E-mail: I.L.Zilberberg@catalysis.ru

Although the mechanism of selective oxidation is commonly considered on base of stepwise Mars-van-Krevelen reaction scheme, the associative mechanism might also take place as was shown for the formaldehyde-to-formic-acid oxidation with the $V_2O_5/TiO_2$ catalyst where the oxidative elimination of oxidized substrate was suggested [15]. Mars-van-Krevelen mechanism implies the creation of oxygen vacancies which are to be reoxidized by gaseous oxygen at later reaction stages. The mentioned associated mechanism also includes "lattice-oxygen" migration toward formaldehyde to form surface formiate species.

The energy of the oxygen abstraction from the surface (most probably from vanadyl groups) is a key factor of activity for supported vanadia [16]. Oxygen vacancies may appear from three oxygen position: a) terminal oxygen, e.g. vanadyl oxygen in case of $V_2O_5$; b) bridged twofold or threefold position; c) interfaced V-O-Ti position as in case of $V_2O_5/TiO_2$.

By means of the IR spectroscopy Trifiró et al [17] showed that partial oxidation catalysts (e.g. $V_2O_5$, $MoO_3$) contain terminal oxygen centers unlike the total oxidation catalysts (e.g. $MnO_2$, $Co_3O_4$, $NiO$, $CuO$, $Fe_2O_3$).

The calculated energy of abstraction of the terminal oxygen atom from crystal $V_2O_5$ obtained in different plane-wave DFT models ranges from 4.83 eV (PW91/VASP [18]), 5.08 eV (PW91/VASP [19]), 6.47 eV (DeMon [20]) and (PBE/PWSCF [21]). All obtained values seem to be too high to explain high activity of supported vanadia catalysts. In particular, $V_2O_5/TiO_2$ are high performance catalysts for the selective oxidation with the rate of the oxidation reactions varying by 3 orders of magnitude depending on the support [22]:

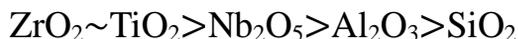

$$ZrO_2 \sim TiO_2 > Nb_2O_5 > Al_2O_3 > SiO_2$$

Analogous data was presented by Wachs [23]:

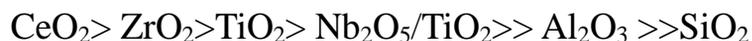

$$CeO_2 > ZrO_2 > TiO_2 > Nb_2O_5/TiO_2 >> Al_2O_3 >> SiO_2$$

One of the possible explanations of this paradox is that the detachment of an oxygen atom from the surface of a particular catalyst leads to the formation of additional bonds resulting in an appreciable mobility of oxygen. Sauer and coauthors [19] have theoretically shown possibility of such compensation due to the formation of the interlayer V-O-V bond on a catalyst surface consisting of two $V_2O_5$ layers. (Figure 1) Due to this compensation, the detachment energy of the oxygen atom decreases to 80 kcal/mol.

It also well-known that the most active catalysts are those with the supports close to monolayer. This fact points to the extremely strong interaction between vanadia and the support, in particular, titania. Despite numerous studies, the origin of the effect of the latter oxide (in this case, Ti) on the activity of transition metal oxide ($V_2O_5$) has not been elucidated yet. Two basic mechanisms of the support effect have been discussed in literature:

- formation of a chemical bond between the catalyst metal and the support metal through oxygen

- support effect, as a structural factor, due to the stabilization of a catalytically active metal in a form differing that for a bulk oxide.

Li and Altman [24] have demonstrated that no bond between V-O-Ti and the support is required for vanadia to exhibit activity in oxidative dehydration. It has also been shown that dehydration of 1-propanol, yielding propanal, is present both on premonolayer and multilayer supports without considerable changes in activation energy; while a poorly ordered multilayer support of titania with vanadia, as well as the bulk vanadia, are inactive in this reaction [25,26]. These results attest to the fact that TiO$_2$ support enhances activity of vanadia due to the stabilization of the structure of active particles.

Witko et al. [18] have considered the formation of oxygen vacancies on different low-index faces (100), (010) and (001) of V$_2$O$_5$ with different vacancy concentrations. Since oxygen of three types is present in V$_2$O$_5$; one-fold O(1), two-fold O(2), three-fold O(3), respectively; the formation of three vacancy types is possible. The lowest energy of oxygen vacancy formation was shown to be 79.8 kcal/mol (3.46 eV) and 73.6 kcal/mol (3.19 eV) for O(1) and O(2) vacancies on the (100) face of V$_2$O$_5$, respectively. On the most energetically stable (001) surface, the lowest energy of vacancy formation corresponds to the formation of O(1) vacancies (vanadyl oxygen), which is due to the higher relaxation energy associated with the formation of a bond between the V atom after the vanadyl oxygen is removed, and the vanadyl oxygen atom of the next layer.

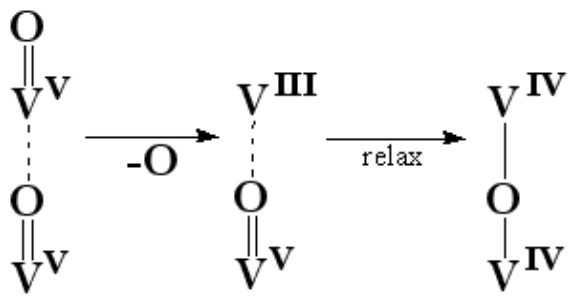

Figure 1: Formation of interlayer V-O-V bonds with the compensation of the separation energy of a vanadyl oxygen atom. Picture is reproduced from article of Ganduglia-Pirovano and Sauer[19].

Minot et al. [27] suggested that the activity of various oxygen center on the surface of supported vanadia might be revealed in the adsorption energy of the atomic hydrogen which is known as a strong reducing agent.

In this paper, it is suggested that the analogous compensation is possible due to the formation of V-O-V bonds not only between the neighboring layers of vanadia, but also within one layer upon the migration of the vanadyl oxygen toward reduced vanadium center to form a bridge bonding. This effect can reveal substantial compensation effect even for a single layer of vanadia on supported surface. Such compensation of the V=O bond dissociation has been predicted for various supports and different coverage rates for any support surface. In the support-free model, the change in distance between vanadyl centers makes it possible to simulate both various supports and coverage rates of the support with vanadia.

## II. Methods and Models

The calculations were performed with the GAUSSIAN98 program package [28] using DFT technique [29]. The hybrid exchange functional Becke [30,31] combined with the Lee-Yang-Parr [32] correlation functional B3LYP (UB3LYP, for the open-shell singlet and triplet states) was used for all treated structures. The 6-311G** basis set was used for all atoms.[33]

An earlier described cluster model of vanadia dimer supported on titania [21] was adopted in present study. The $TiO_2$ support was replaced by terminal hydrogen atoms to have a possibility to vary the V-V distance without any restrictions. Despite the absence of the support, the dissociation energy of the vanadyl bond is surprisingly well described by this model. To prove this, the vanadyl bond dissociation energy in the model containing titanium support (Figure 2, **A**) was calculated versus the model with terminal hydrogens atoms (Figure 2, **C**). Both models were optimized. In the model with hydrogens (Figure 2) terminal OH groups were fixed. The energy of dissociation of the vanadyl bond with the titania support was 140 kcal/mol, which perfectly agrees with that energy of 142.74 kcal/mol calculated using the support-free model.

The vanadyl bond dissociation energy was defined as the energy difference between the sum of energies of the reduced form and the oxygen atom the oxidized form: $D_0(V=O)= E_{Red}+E_{tot}(O) - E_{Ox}$. For each value of distance d which was varied from 3.0 Å to 5.0 Å with the a step 0.2 Å the total energy of the system in oxidized form (Figure 3,**E**) and reduced form (Figure 3,**F**) was calculated.

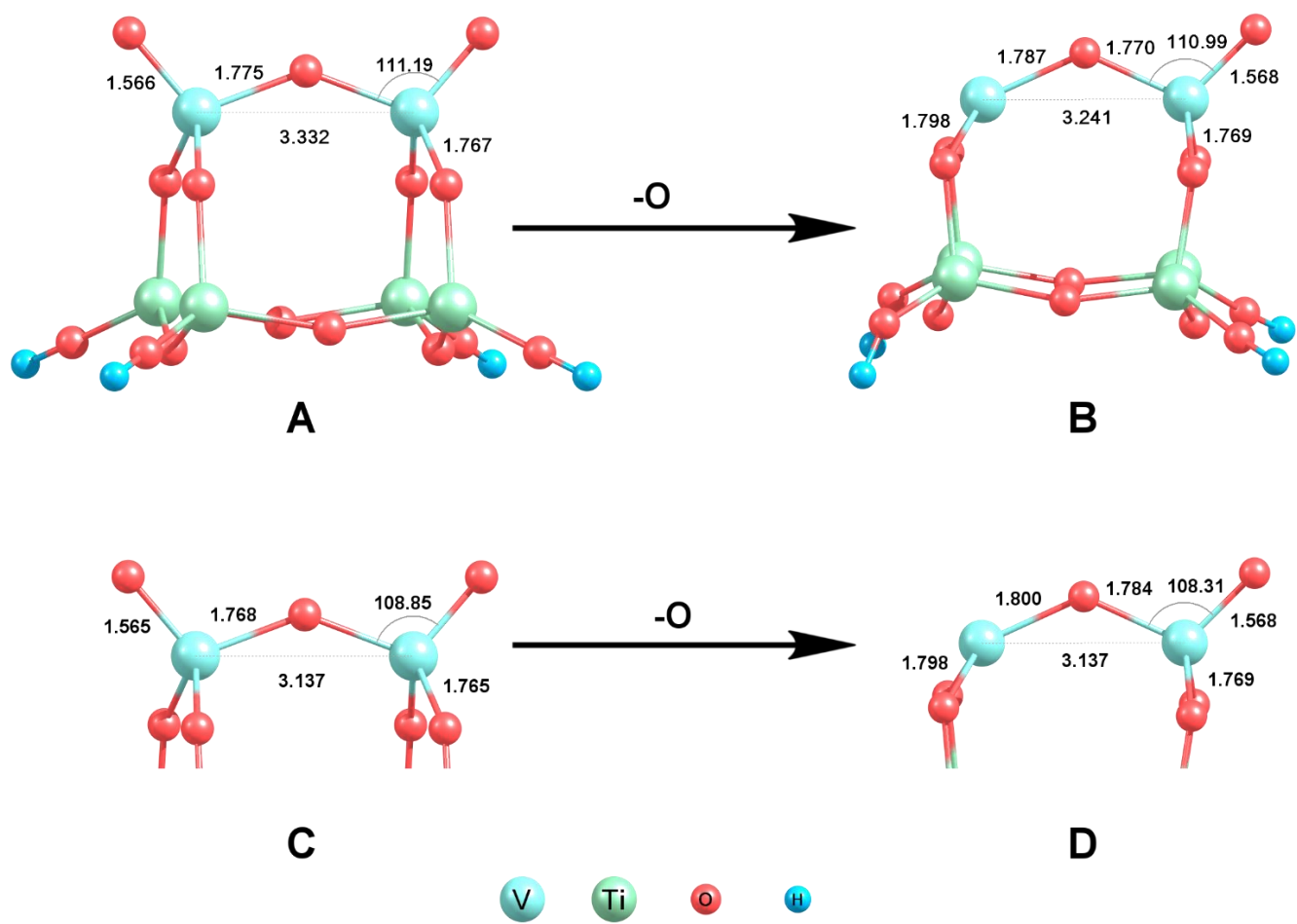

Figure 2: Cluster models of TiO2-supported V2O5 (A) and its reduced form (B). The model for the $V_2O_5$ dimer without support (imitated by hydrogen atoms) in initial oxydized (C) and reduced (D) forms. The numerals show the interatomic distances (Å) and ∠O-V-O.

To determine the energy of reduced system as a function of distance **d**, we used two dimers, in one of which the vanadyl oxygen atom being removed. Thus, one of the dimers is a model of the reduced $V_2O_4$ (Figure 3,**G**).

## III. Results and Discussion

There are two ways of the formation of surface vanadium oxide particles: replacement of Ti atoms by V atoms on $TiO_2$, bond formation between V and oxygen of TiO2. In the former case, the monolayer cover corresponds to the vanadium content of 5.72 V/nm$^2$; in the latter case, it corresponds to 11.44 V/nm$^2$.

When two oxidized clusters are brought together closer than d=4.0 Å, vanadyl oxygen atoms repulse from each other (Figure 3) deviating from the V-O-V plane at the dihedral angle 70.8° at d=3.0 Å. This repulsion is though relatively small from energetic point of view increasing the total energy of the system by only 10 kcal/mol. At distances $d \geq 4$ Å repulsion of this type quickly disappears (Figure 3,**E**), resulting in the decrease of mentioned dihedral angle down to 4° and less.

Two distinct regions show up on the dependence of the dissociation energy on the distance $d$. One of them, 3 Å≤d ≤ 4 Å, is associated with the formation of the V-O-V bridge between the vanadate particles. In this bridge the V-O bond length varies from 1.78 Å at $d = 3$ Å to 2.1 Å at $d = 4$ Å (Figure 3). Another region is associated with the formation of the "non-perturbed" vanadyl bond V=O with the bond length of 1.57 Å. The transition between these two regions occurs jumpwise at $d = 4$ Å.

At $d ≤ 4.0$Å, the formation of the bridge oxygen atom coming from the vanadyl oxygen atom (Figure 3,**G**) occurs, making it possible to equalize the oxidation states of the vanadium centers connected by bridge oxygen atom. In other words, the process V$^{5+}$=O + V$^{3+}$ → V$^{4+}$-O-V$^{4+}$ takes place (Figure 3,**G**). The formation of the bridge oxygen atom results in a considerable decrease in the dissociation energy to 35 kcal/mol at $d = 3$ Å. At these distances, there is one unpaired electron on V2 and V2' atoms, which can be seen from data listed in Table S1. The value <S$^2$> at d ≤ 4.0 almost constant and amounts to 1.01 (Figure 4).

As seen from data represented, due to the formation of the V-O-V bridge bond between the monomers, the abstraction energy of the oxygen atom decreases from 140 kcal/mol (at distance of 5.0 Å, which is equivalent to single V$_2$O$_5$) to 35 kcal/mol (which is equivalent to the formation of polyvanadyl structures) (Figure 4).

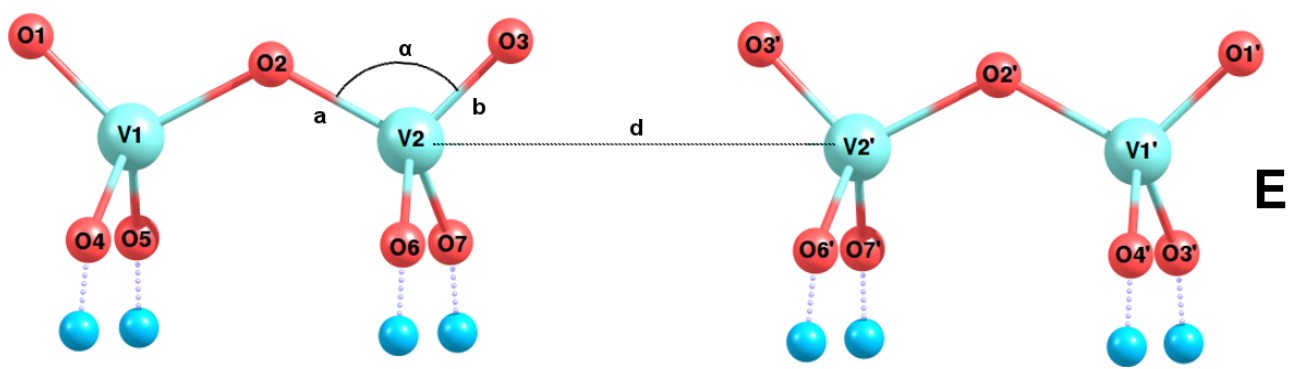

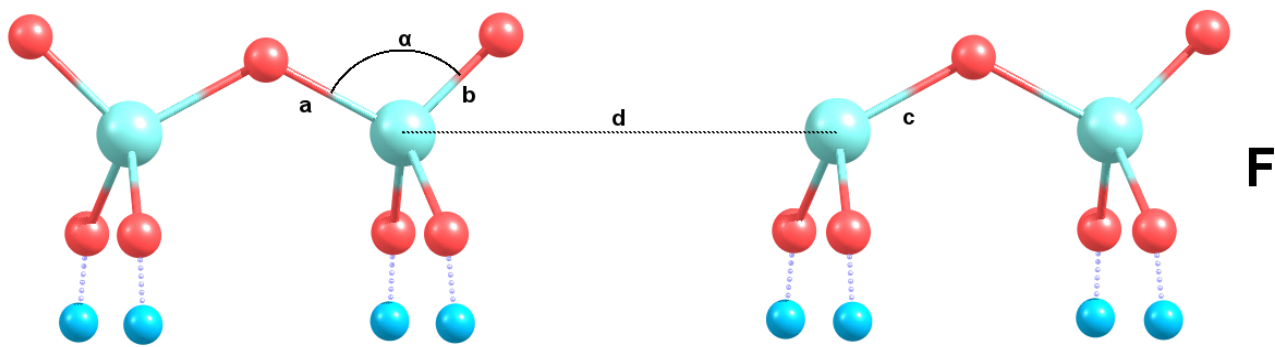

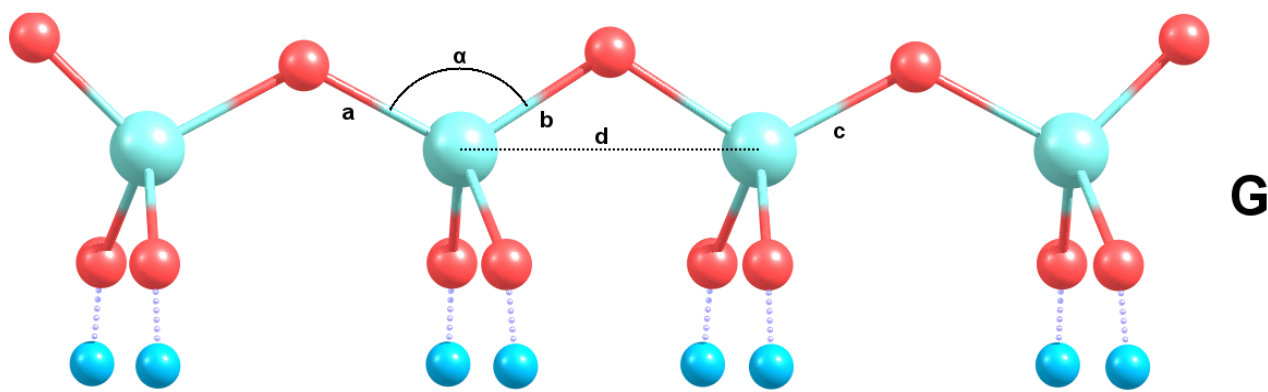

| Structure | d, Å | α | a, Å | b, Å | c, Å |
|-----------|------|-----|-------|-------|-------|
| E | 5 | 107 | 1.771 | 1.559 | 1.772 |
| F | 5 | 110 | 1.765 | 1.566 | 1.805 |
| G | 3 | 120 | 1.769 | 1.773 | 1.769 |

Figure 3: Formation of the bridge oxygen upon the vanadyl oxygen abstraction at different distances between dimers.

The dissociation energy abruptly changes from 103 kcal/mol at d = 3.8 Å to 129 kcal/mol at d = 4.0 Å. The edge vanadium atoms (V1 и V1') and the corresponding vanadyl oxygen atoms (O1 и O1') do not change their positions, in particularly, the distances V1=O1 and V1'=O1' and angles O1=V1-O2 and O1'=V1'-O2'. Thus, the data can also be used for different V1-V2 distances in the dimer, and, therefore, for different supports.

$$D_0 (V = O) = E_{Red} + E_{tot}(O) - E_{Ox}, \text{ where } E_{tot}(O) = -75.08539 \text{ a.u.} \tag{1}$$

Table 1. The $S_z=0$ solutions for two dimers (Figure 2) upon the the vanadyl oxygen abstraction at various distances d(V-V) between vanadyl groups: the mean value for squared spin operator $<S^2>$, distances between interacting vanadium centers V2 and V2' and bridge oxygen atom O3 (V2-O3), d(V2'-O3), and the dissociation energy of vanadyl bond $D_0(V=O)$.

| d (V-V), Å | $<S^2>$ | d(V2-O3), Å | d(V2'-O3), Å | $D_0(V=O)$, kcal/mol |
|---|---|---|---|---|
| 3.0 | 1.0068 | 1.773 | 1.776 | 35.65 |
| 3.2 | 1.0094 | 1.804 | 1.804 | 43.82 |
| 3.4 | 1.0107 | 1.847 | 1.847 | 58.23 |
| 3.6 | 1.0121 | 1.902 | 1.912 | 78.62 |
| 3.8 | 1.0137 | 1.967 | 1.998 | 103.23 |
| 4.0 | 1.0159 | 1.955 | 2.183 | 128.80 |
| 4.2 | 0.8728 | 1.581 | 3.13 | 129.73 |
| 4.4 | 0.8587 | 1.575 | 3.388 | 134.84 |
| 4.6 | 0.8521 | 1.573 | 3.581 | 138.58 |
| 4.8 | 0.8474 | 1.571 | 3.786 | 140.97 |
| 5.0 | 0.8423 | 1.567 | 3.985 | 142.46 |

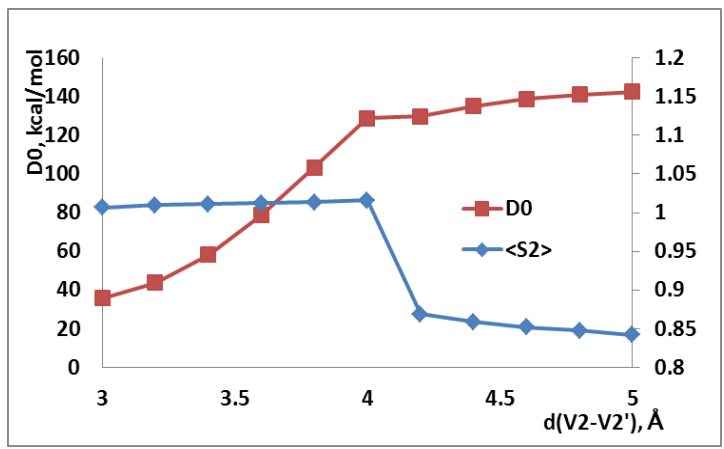

Figure 4: The dissociation energy (D₀) of the vanadyl group and <S²> depending on the interdimer distance d(V2-V2').

When distance d decreases down to 4 Å, the dissociation energy and <S²> experience a break corresponding to equalization of vanadium oxidation states: $V^{5+}=O\cdots V^{3+} \rightarrow V^{4+}-O-V^{4+}$. This abrupt change results from the bridge oxygen appearance. The expected values of <S²> jumps from 0.85 to 1.02 corresponding to spin transfer from $V^{3+}$ center to neighboring $V^{5+}$ center (Figure 5).

As seen from Table S1, at d < 4 Å, the non-zero (positive and negative) spin density takes place only on those vanadium atoms (V2 и V2') that participate in the formation of the V-O-V bridge. At d > 4 Å the spin density on all vanadium atoms is zero that can be described assigned to as the $V^{5+}=O\cdots V^{3+}$ configuration having antiparallel spins on different $d$ orbitals of the $V^{3+}$ center. Despite this fact, <S²> ≃ 0.85, which is attributed to the fact that on vanadium atom V2 that was reduced by the elimination of the oxygen atom, there are two electrons, α and β, located at different atomic orbitals. Figure 4 shows the unpaired electron distribution at different $d$.

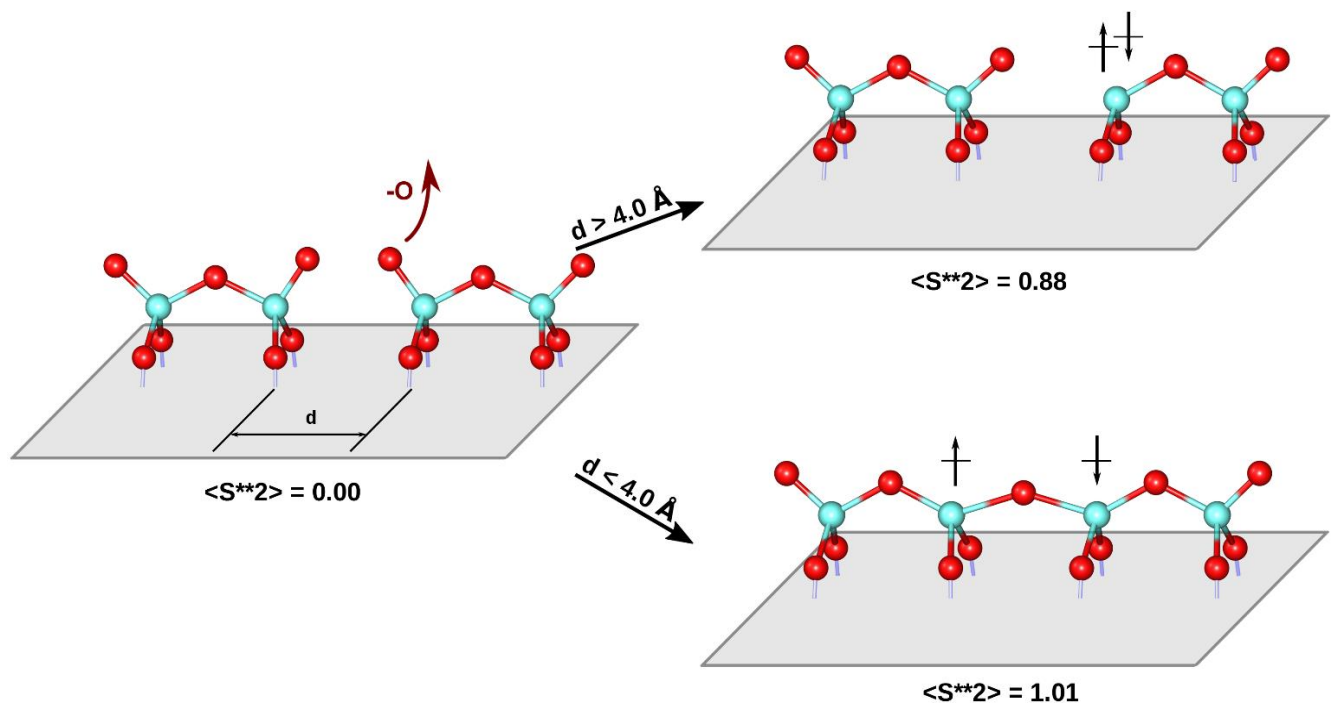

Figure 5. Scheme of the interaction between two vanadyl groups upon the abstraction of oxygen from one of them at various distances between vanadium centers.

The bridge oxygen formation appreciably changes the geometrical parameters of dimers near the vanadyl oxygen atom (Table 1 and Table S1): if the distance between vanadium centers exceeds ca. 4 Å, then adjacent vanadyl oxygen moves into a bridge position.

Thus, for instance, when the distance rises up 4.0 Å and more, the formation of V2=O3 vanadyl bond is preferred over the formation of the bridge oxygen atom V2O3V2'. When $d$ decreases to 4.0 Å, an abrupt change in O3V2' occurs, indicating the preference of formation of one vanadyl bond V=O.

Above described results agree fairly well with the study by Avdeev and Tapilin who obtained ΔH of 5.74 eV (132.3 kcal/mol) for the reaction

O=V$^{5+}$–O–V$^{5+}$=O + O=V$^{5+}$–O–V$^{5+}$=O → O=V$^{5+}$–O–V$^{5+}$=O + V$^{3+}$–O–V$^{5+}$=O + O

for d=4.28Å using periodic DFT approach [21]. For this distance, almost the same energy of 131.8 kcal/mol is given by the present model.

# IV. Conclusions

In this work, the effect of compensation of the vanadyl group dissociation via formation of the bridge oxygen from neighboring vanadate particle has been predicted. The dissociation energy of V=O achieves its minimal value of 36 kcal/mol at separation between vanadate particles of 3 Å.

On base of obtained results one may conclude that the geometry of supporting surface (more specifically, the minimal distance between surface cations which couple the vanadate particles) is responsible for the dissociation energy of the vanadyl group and so mobility of vanadyl oxygen in case of low-density (less than monolayer) vanadate particles on this surface. For each fixed coverage on a given support, there is a particular discrete spectrum of distances between separated vanadate particles, corresponding to a discrete spectrum of the V=O dissociation energies.

# V. Acknowledgements

This study was supported by Russian Academy of Sciences and Federal Agency for Scientific Organizations. Authors thank the Siberian Supercomputer Center for a generous allotment of computer time.

# Supporting information

| d(V-V), Å | Mulliken atomic spin density | | | | | | | | |
|---|---|---|---|---|---|---|---|---|---|
| | O1 | V1 | O2 | V2 | O3 | O1' | V1' | O2' | V2' |
| 3 | -0.002 | 0.0018 | -0.027 | 0.947 | 0 | 0.002 | -0.018 | 0.028 | -0.947 |
| 3.2 | -0.001 | 0.014 | -0.028 | 0.956 | 0 | 0.001 | -0.014 | 0.028 | -0.956 |
| 3.4 | -0.001 | 0.011 | -0.029 | 0.967 | 0 | 0.001 | -0.011 | 0.029 | -0.967 |
| 3.6 | -0.001 | 0.008 | -0.029 | 0.978 | -0.001 | 0.001 | -0.008 | 0.03 | -0.979 |
| 3.8 | -0.001 | 0.007 | -0.03 | 0.99 | -0.002 | 0 | -0.006 | 0.031 | -0.992 |
| 4 | -0.001 | 0.007 | -0.028 | 0.997 | -0.016 | 0.001 | -0.005 | 0.035 | -1.007 |
| 4.2 | 0 | 0 | 0.001 | -0.038 | 0.01 | 0.002 | -0.043 | -0.017 | 0.035 |
| 4.4 | 0 | 0 | 0 | 0.017 | -0.003 | -0.002 | 0.04 | 0.018 | -0.019 |
| 4.6 | 0 | 0 | 0 | -0.011 | 0.002 | 0.002 | -0.039 | -0.017 | 0.014 |
| 4.8 | 0 | 0 | 0 | 0.008 | -0.001 | -0.002 | 0.037 | 0.017 | -0.011 |
| 5 | 0 | 0 | 0 | -0.013 | 0.007 | 0.003 | -0.039 | -0.016 | 0.009 |

Table S1: Mulliken atomic spin density as a function of interdimer distance d for the reduced dimer system.

| Element | Coordinates | | |
|---|---|---|---|
| | X | Y | Z |
| H | -4.859195000 | 2.104169000 | -1.425803000 |
| H | -4.858572000 | 1.672953000 | 1.903789000 |
| H | -1.282863000 | 2.105420000 | -1.426310000 |
| H | -1.282240000 | 1.674204000 | 1.903282000 |
| H | 1.277739000 | 2.106314000 | -1.426677000 |
| H | 1.278362000 | 1.675097000 | 1.902922000 |
| H | 4.854058000 | 2.107563000 | -1.427181000 |
| H | 4.854681000 | 1.676347000 | 1.902412000 |
| O | -4.830952000 | 1.085409000 | -1.234162000 |
| O | -4.830449000 | 0.736626000 | 1.458934000 |
| O | -1.310365000 | 1.086640000 | -1.234662000 |
| O | -1.309861000 | 0.737857000 | 1.458435000 |
| O | 1.305976000 | 1.087553000 | -1.235028000 |
| O | 1.306480000 | 0.738771000 | 1.458059000 |
| O | 4.826583000 | 1.088783000 | -1.235532000 |
| O | 4.827087000 | 0.740000000 | 1.457564000 |
| O | -5.760410000 | -1.270992000 | -0.170110000 |
| O | -3.057070000 | -0.979278000 | -0.132408000 |
| O | 0.000367000 | -1.183606000 | -0.901161000 |
| O | -0.001178000 | -1.372967000 | 0.554108000 |
| O | 3.054506000 | -0.977433000 | -0.131842000 |
| O | 5.758066000 | -1.266914000 | -0.171668000 |
| V | -4.638491000 | -0.190149000 | -0.030262000 |
| V | -1.501565000 | -0.189059000 | -0.030708000 |
| V | 1.498435000 | -0.188008000 | -0.031133000 |
| V | 4.635358000 | -0.186910000 | -0.031578000 |

Table 2: Coordinates of Oxidized structure with d(V-V)=3.0 Å

| Element | Coordinates | | |
|---------|-------------|---|---|
|         | X           | Y | Z |
| H | -4.859195000 | 2.104169000 | -1.425803000 |
| H | -4.858572000 | 1.672953000 | 1.903789000 |
| H | -1.282863000 | 2.105420000 | -1.426310000 |
| H | -1.282240000 | 1.674204000 | 1.903282000 |
| H | 1.277739000 | 2.106314000 | -1.426677000 |
| H | 1.278362000 | 1.675097000 | 1.902922000 |
| H | 4.854058000 | 2.107563000 | -1.427181000 |
| H | 4.854681000 | 1.676347000 | 1.902412000 |
| O | -4.830952000 | 1.085409000 | -1.234162000 |
| O | -4.830449000 | 0.736626000 | 1.458934000 |
| O | -1.310365000 | 1.086640000 | -1.234662000 |
| O | -1.309861000 | 0.737857000 | 1.458435000 |
| O | 1.305976000 | 1.087553000 | -1.235028000 |
| O | 1.306480000 | 0.738771000 | 1.458059000 |
| O | 4.826583000 | 1.088783000 | -1.235532000 |
| O | 4.827087000 | 0.740000000 | 1.457564000 |
| O | -5.770960000 | -1.262483000 | -0.168677000 |
| O | -3.073581000 | -1.002057000 | -0.137158000 |
| O | -0.001661000 | -1.127568000 | -0.152499000 |
| O | 3.070606000 | -0.999942000 | -0.135267000 |
| O | 5.768284000 | -1.258694000 | -0.170780000 |
| V | -4.638491000 | -0.190149000 | -0.030262000 |
| V | -1.501565000 | -0.189059000 | -0.030708000 |
| V | 1.498435000 | -0.188008000 | -0.031133000 |
| V | 4.635358000 | -0.186910000 | -0.031578000 |

Table 3: Coordinates of Reduced structure with d(V-V)=3.0 Å

| Element | Coordinates | | |
|---------|-------------|-------------|-------------|
|         | X           | Y           | Z           |
| H | -5.359195000 | 2.103994000 | -1.425732000 |
| H | -5.358572000 | 1.672778000 | 1.903860000 |
| H | -1.782863000 | 2.105245000 | -1.426239000 |
| H | -1.782240000 | 1.674029000 | 1.903353000 |
| H | 1.777739000 | 2.106489000 | -1.426748000 |
| H | 1.778362000 | 1.675272000 | 1.902851000 |
| H | 5.354058000 | 2.107738000 | -1.427252000 |
| H | 5.354681000 | 1.676522000 | 1.902341000 |
| O | -5.330952000 | 1.085234000 | -1.234091000 |
| O | -5.330449000 | 0.736451000 | 1.459005000 |
| O | -1.810365000 | 1.086465000 | -1.234591000 |
| O | -1.809861000 | 0.737682000 | 1.458506000 |
| O | 1.805976000 | 1.087728000 | -1.235099000 |
| O | 1.806480000 | 0.738946000 | 1.457988000 |
| O | 5.326583000 | 1.088958000 | -1.235603000 |
| O | 5.327087000 | 0.740175000 | 1.457493000 |
| O | -6.266199000 | -1.263768000 | -0.169025000 |
| O | -3.561760000 | -0.997214000 | -0.138899000 |
| O | -0.874343000 | -1.272988000 | -0.136060000 |
| O | 1.250740000 | -1.715318000 | -0.570900000 |
| O | 3.587008000 | -0.982241000 | -0.028091000 |
| O | 6.261653000 | -1.267468000 | -0.184109000 |
| V | -5.138491000 | -0.190324000 | -0.030191000 |
| V | -2.001565000 | -0.189234000 | -0.030637000 |
| V | 1.998435000 | -0.187833000 | -0.031204000 |
| V | 5.135358000 | -0.186735000 | -0.031649000 |

Table 4: Coordinates of Oxidized structure with d(V-V)=4.0 Å

| Element | Coordinates | | |
|---|---|---|---|
| | X | Y | Z |
| H | -5.359195000 | 2.103994000 | -1.425732000 |
| H | -5.358572000 | 1.672778000 | 1.903860000 |
| H | -1.782863000 | 2.105245000 | -1.426239000 |
| H | -1.782240000 | 1.674029000 | 1.903353000 |
| H | 1.777739000 | 2.106489000 | -1.426748000 |
| H | 1.778362000 | 1.675272000 | 1.902851000 |
| H | 5.354058000 | 2.107738000 | -1.427252000 |
| H | 5.354681000 | 1.676522000 | 1.902341000 |
| O | -5.330952000 | 1.085234000 | -1.234091000 |
| O | -5.330449000 | 0.736451000 | 1.459005000 |
| O | -1.810365000 | 1.086465000 | -1.234591000 |
| O | -1.809861000 | 0.737682000 | 1.458506000 |
| O | 1.805976000 | 1.087728000 | -1.235099000 |
| O | 1.806480000 | 0.738946000 | 1.457988000 |
| O | 5.326583000 | 1.088958000 | -1.235603000 |
| O | 5.327087000 | 0.740175000 | 1.457493000 |
| O | -6.296070000 | -1.243116000 | -0.166459000 |
| O | -3.607334000 | -1.054938000 | -0.141616000 |
| O | 0.784938000 | -1.196932000 | -0.160441000 |
| O | 3.557232000 | -1.008751000 | -0.135363000 |
| O | 6.267911000 | -1.254325000 | -0.170549000 |
| V | -5.138491000 | -0.190324000 | -0.030191000 |
| V | -2.001565000 | -0.189234000 | -0.030637000 |
| V | 1.998435000 | -0.187833000 | -0.031204000 |
| V | 5.135358000 | -0.186735000 | -0.031649000 |

Table 5: Coordinates of Reduced structure with d(V-V)=4.0 Å

| Element | Coordinates | | |
|---|---|---|---|
| | X | Y | Z |
| H | -5.859195000 | 2.103819000 | -1.425661000 |
| H | -5.858572000 | 1.672603000 | 1.903931000 |
| H | -2.282863000 | 2.105070000 | -1.426168000 |
| H | -2.282240000 | 1.673854000 | 1.903424000 |
| H | 2.277739000 | 2.106664000 | -1.426819000 |
| H | 2.278362000 | 1.675447000 | 1.902780000 |
| H | 5.854058000 | 2.107913000 | -1.427323000 |
| H | 5.854681000 | 1.676697000 | 1.902270000 |
| O | -5.830952000 | 1.085059000 | -1.234020000 |
| O | -5.830449000 | 0.736276000 | 1.459076000 |
| O | -2.310365000 | 1.086290000 | -1.234520000 |
| O | -2.309861000 | 0.737507000 | 1.458577000 |
| O | 2.305976000 | 1.087903000 | -1.235170000 |
| O | 2.306480000 | 0.739121000 | 1.457917000 |
| O | 5.826583000 | 1.089133000 | -1.235674000 |
| O | 5.827087000 | 0.740350000 | 1.457422000 |
| O | -6.760732000 | -1.272901000 | -0.169143000 |
| O | -4.068107000 | -0.984895000 | -0.138628000 |
| O | -1.189122000 | -1.392699000 | -0.189821000 |
| O | 1.376451000 | -1.264725000 | -0.167824000 |
| O | 4.068269000 | -0.999324000 | -0.138561000 |
| O | 6.767349000 | -1.257466000 | -0.170189000 |
| V | -5.638491000 | -0.190499000 | -0.030120000 |
| V | -2.501565000 | -0.189409000 | -0.030566000 |
| V | 2.498435000 | -0.187658000 | -0.031275000 |
| V | 5.635358000 | -0.186560000 | -0.031720000 |

Table 6: Coordinates of Oxidized structure with d(V-V)=5.0 Å

| Element | Coordinates | | |
|---------|-------------|-------------|-------------|
|         | X           | Y           | Z           |
| H | -5.859195000 | 2.103819000 | -1.425661000 |
| H | -5.858572000 | 1.672603000 | 1.903931000 |
| H | -2.282863000 | 2.105070000 | -1.426168000 |
| H | -2.282240000 | 1.673854000 | 1.903424000 |
| H | 2.277739000 | 2.106664000 | -1.426819000 |
| H | 2.278362000 | 1.675447000 | 1.902780000 |
| H | 5.854058000 | 2.107913000 | -1.427323000 |
| H | 5.854681000 | 1.676697000 | 1.902270000 |
| O | -5.830952000 | 1.085059000 | -1.234020000 |
| O | -5.830449000 | 0.736276000 | 1.459076000 |
| O | -2.310365000 | 1.086290000 | -1.234520000 |
| O | -2.309861000 | 0.737507000 | 1.458577000 |
| O | 2.305976000 | 1.087903000 | -1.235170000 |
| O | 2.306480000 | 0.739121000 | 1.457917000 |
| O | 5.826583000 | 1.089133000 | -1.235674000 |
| O | 5.827087000 | 0.740350000 | 1.457422000 |
| O | -6.790926000 | -1.246571000 | -0.166895000 |
| O | -4.095765000 | -1.036311000 | -0.138886000 |
| O | 1.346818000 | -1.243538000 | -0.167222000 |
| O | 4.062625000 | -1.001193000 | -0.136025000 |
| O | 6.766200000 | -1.257288000 | -0.170726000 |
| V | -5.638491000 | -0.190499000 | -0.030120000 |
| V | -2.501565000 | -0.189409000 | -0.030566000 |
| V | 2.498435000 | -0.187658000 | -0.031275000 |
| V | 5.635358000 | -0.186560000 | -0.031720000 |

Table 7: Coordinates of Reduced structure with d(V-V)=5.0 Å